\documentclass{article}
\usepackage{spconf,amsmath,graphicx}
\usepackage{colortbl}
\usepackage{amsmath}
\usepackage{hyperref}
\definecolor{Gray}{gray}{0.9}

\title{End-to-end deep multi-score model for No-reference stereoscopic image quality assessment}

\begin{document}
%
\maketitle
\vspace{-3mm}
\begin{abstract}

Deep learning-based quality metrics have recently given significant improvement in Image Quality Assessment (IQA). In the field of stereoscopic vision, information is evenly distributed with slight disparity to the left and right eyes. However, due to asymmetric distortion, the objective quality ratings for the left and right images would differ, necessitating the learning of unique quality indicators for each view. Unlike existing stereoscopic IQA measures which focus mainly on estimating a global human score, we suggest incorporating left, right, and stereoscopic objective scores to extract the corresponding properties of each view, and so forth estimating stereoscopic image quality without reference. Therefore, we use a deep multi-score Convolutional Neural Network (CNN). Our model has been trained to perform four tasks: First, predict the left view's quality. Second, predict the quality of the left view. Third and fourth, predict the quality of the stereo view and global quality, respectively, with the global score serving as the ultimate quality. Experiments are conducted on Waterloo IVC 3D Phase 1 and Phase 2 databases. The results obtained show the superiority of our method when comparing with those of the state-of-the-art. \textit{The implementation code can be found at:  \href{https://github.com/o-messai/multi-score-SIQA}{https://github.com/o-messai/multi-score-SIQA}}
\end{abstract}

\vspace{-3mm}

\section{Introduction}
\label{sec:intro}
\vspace{-3mm}

Stereoscopic images are now widely used in a variety of applications including 3D medical imaging, virtual reality, and 3D video games \cite{messai2020adaboost}. Several processes are typically conducted to such images (compression, transmission, restoration, etc.), each of which can impact the perceived quality \cite{chetouani2018ICIP,Chetouani20EUSIPCO}. This problem has prompted the computer vision field to develop sophisticated quality measurements that forecast the perceived impact of these distortions, known as Stereoscopic Image Quality Assessment (SIQA). There are two types of SIQA methods: subjective SIQA and objective SIQA. The former approaches are costly and time consuming because they rely on human score opinion to judge quality, whereas the latter are inexpensive and quick since they rely on machine algorithmic score. However, because humans are the final recipients of 3D content, it is important to validate the metric output with a subjective evaluation, namely human visual quality assessment. It is mostly expressed in terms of Mean Opinion Score (MOS) or Difference Mean Opinion Score (DMOS). 

The need for SIQA development first gained traction in 2009, when Benoit \textit{et al.} \cite{benoit2009quality} suggested a methodology that combines two metrics. They begin by calculating the difference between the left and right reference images and the corresponding distorted images. The difference between the pure stereo image disparity map and the distorted ones is then computed. Another remarkable FR-SIQA \cite{chen2013full} metric has been presented that outperforms the latter. The authors used a linear formulation of cyclopean view that is impacted by binocular competition between left and right views. Ma \textit{et al.} \cite{ma2017full} also presented NR-SIQA, which employs binocular combination idea to conduct Human Visual System (HVS) simulation.

\begin{figure*}[h]
    \centering
    \includegraphics[height=7.0cm]{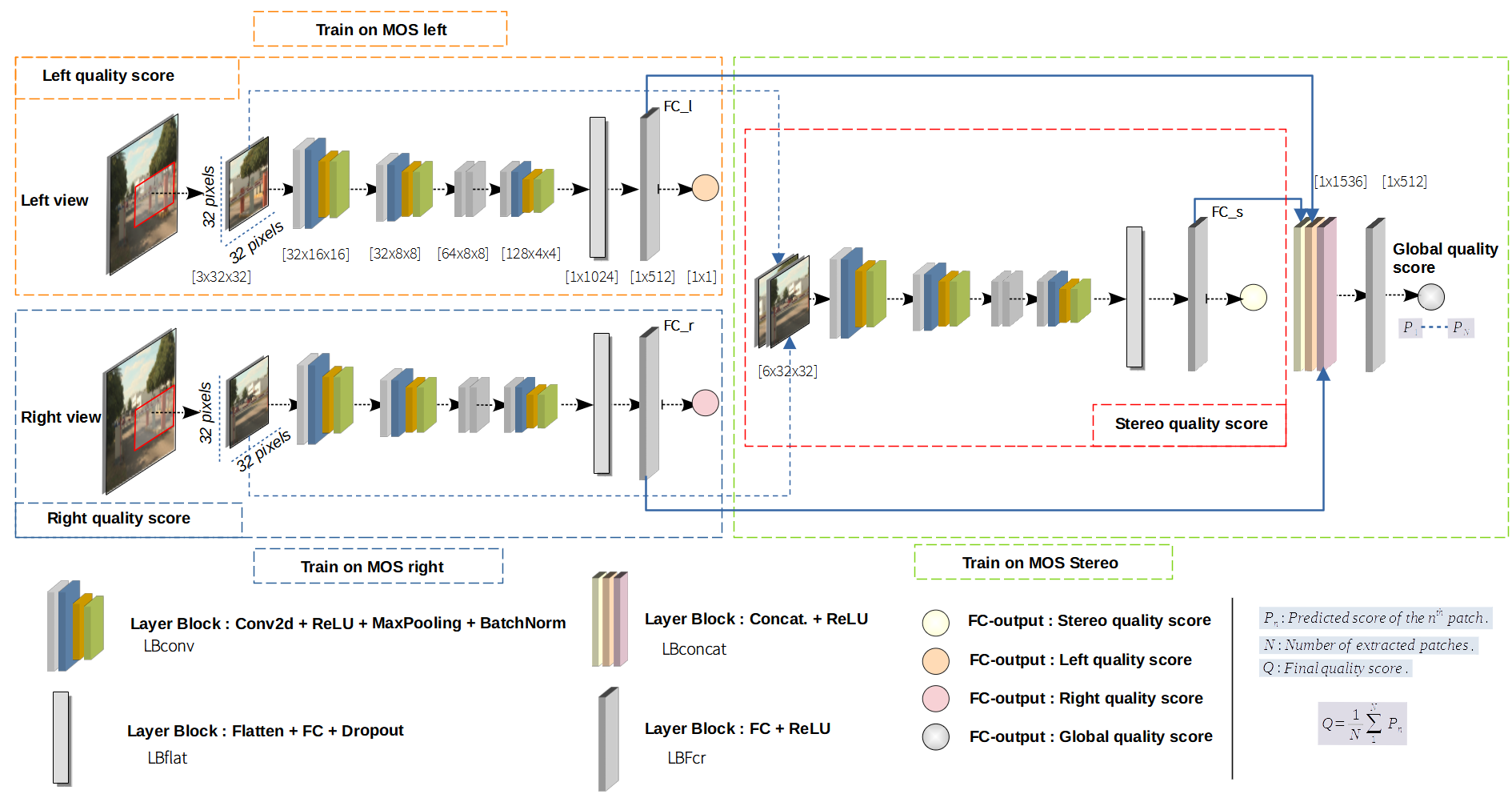}
    \caption{The proposed multi-score CNN based model. The network outputs are: left, right, stereo, and global quality scores.}
    \label{fig:CNN_model}
\end{figure*}

Due to the general benefits they provide, SIQA researchers are increasingly relying on NR measures. Therefore, most present SIQA approaches are dedicated to NR-SIQA to satisfy the requirements of most modern applications. Machine learning approaches, in particular Deep Learning (DL) \cite{Mahmoudi2020ICIP,MAHMOUDI2020PRL}, enabled the automatic extraction of the best features, allowing them to outperform handwritten characteristics. However, in the SIQA domain, the learning techniques may differ from one measure to another, but the success is strongly dependent on the derived quality characteristics. Simulating the HVS's quality assessment behavior during binocular vision is another key aspect. The latter, on the other hand, is still in its early phases. In the following, we briefly address the recent suggested NR-SIQA metrics: In \cite{sun2020learning}, saliency information \cite{Ilyass19NCAA,PR20Ilyass}
was used to select salient and non-salient patches for local feature extraction. The authors used reference stereo images to build local quality maps, which were then used as labels to train deep CNN as a local quality-aware structures predictor. The latter are then combined into a final quality score. Image segmentation techniques have also been used in NR-SIQA methods. In \cite{liu2020no} a superpixel segmentation based on the K-mean clustering technique is used. Then, using a regression model, spatial entropy and Natural Scene Statistics (NSS) features are retrieved from these superpixel regions to generate quality scores. CNNs can be applied to a variety of topics, for instance, authors in \cite{yang2018blind} have modeled the human visual cortex using the deep CNN auto-encoder. The CNN-based auto-encoder is also used in \cite{yang2018predicting} to achieve high-level features, where the authors first compute a gray level cyclopean image, difference and summation image from the input stereoscopic view. Then, series of feature extraction have been conducted from these images. Recently in \cite{wang2021no}, monocular and binocular quality features, including texture and energy features, are first retrieved, and then ensemble learning is utilized to map the quality score. Meanwhile, in \cite{bourbia2021multi} a multi-task CNN model is proposed to extract NSS features as an auxiliary task, and to produce quality score as primary task. Besides auto-encoders, multi-task models are increasingly being deployed in SIQA domain, but none of them explore the concept of multi-score predictions (e.g., right, left and stereo score). Because binocular quality attributes differ from left to right, especially in the situation of asymmetric distortion, that yield binocular rivalry. An independent quality-aware indicator must then be learned from the stereoscopic view. Moreover, many of NR-SIQA approaches require a fixed resolution for the input image. Therefore, in this paper, we propose an NR-SIQA measure that addresses these drawbacks.

The remainder of this paper is organized as follows. In Section \ref{sec:Proposed_method}, we describe the proposed method. Then, we present the experimental results in the Section \ref{sec:experimental}. Finally, we give some concluding remarks in Section \ref{sec:conclusion}.

\vspace{-3mm}
\section{PROPOSED MULTI-SCORE METHOD}
\label{sec:Proposed_method}
\vspace{-3mm}


Fig. \ref{fig:CNN_model} presents the flowchart of the proposed model which combines three sub-networks: for the stereoscopic image input, each view corresponds to sub-network. These two sub-networks are distinct in order to accommodate for independent binocular information and aim to mimic binocular vision. Then, to simulate the pathway of optic nerve in HVS, the left and right views are concatenated as input for the third sub-network. It is worth noting that the proposed metric scheme takes the input RGB (Red, Green, Blue) stereo image without any pre-processing and provides four output scores (e.i., left, right, stereo and global score), whereas most SIQA metrics convert the input images to a typical gray tone as a pre-treatment step for the CNN model and give a single score. As demonstrated in our recent work \cite{messai20223d}, using three channels as input rather than a single gray scale tends to increase the performance by preserving the perceived original distortion effects and judged by the human during the evaluation process. 

\vspace{-3mm}
\subsection{Extraction of Independent Features}
\vspace{-3mm}


Independent binocular features are retrieved through the proposed multi-score prediction model. The left sub-network seeks to extract features from the left view, while the right one seeks to extract features from the right view using left and right objective score, respectively. However, using just left and right sub-networks, we may extract features individually from the left and right images, and then weight-average the score to determine final quality. However, as previously stated, binocular rivalry occurs when the two views of a stereo pair exhibit different types or degrees of distortion. As a result, the average quality of the left and right views cannot predict the objective quality of the most often viewed stereo image. To illustrate this point, we take 55 distorted stereoscopic images. We then compute the average of $MOS_{left}$ and $MOS_{right}$, and compare it to the corresponding $MOS_{stereo}$. The comparison is carried out by simply computing the absolute value of difference $D$ as follows:

\begin{equation}
    D = |\frac{(MOS_{left} + MOS_{right})}{2} - MOS_{stereo}|
\end{equation}

\begin{figure}[h]
    \centering
    \includegraphics[height=3.9cm]{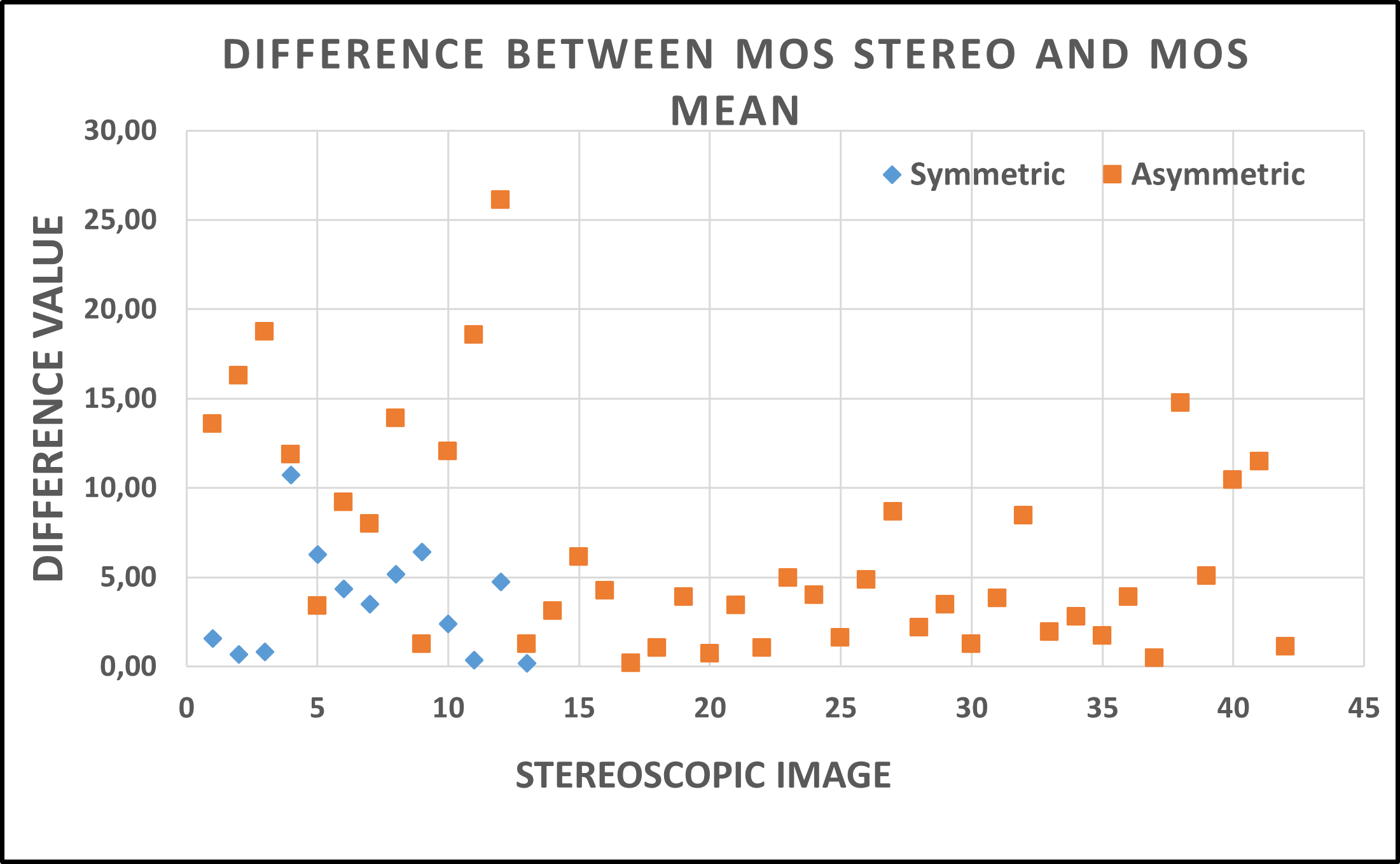}
    \caption{Difference values of $MOS_{stereo}$ with $MOS_{mean}$  mean of 55 distorted views produced from a pristine stereoscopic image called \textit{Art} from Waterloo IVC Phase 1 \cite{wang2014quality}}
    \label{fig:Mos_value}
\end{figure}

The result is illustrated in Fig. \ref{fig:Mos_value}. As can be seen, there is always a mismatch between the stereo quality score and the average left and right score, particularly with asymmetric distortion, where we note up to a 25-point difference in term of MOS. This basic analysis prompted us further more to propose the multi-score prediction system described in the following subsection.

\vspace{-3mm}

\vspace{-2mm}
\subsection{Multi-score prediction}
\vspace{-3mm}


In this study, we present an end-to-end CNN multi-score architecture composed of three sub CNNs, each one contains the following blocks in the correct sequence as shown in Fig. \ref{fig:CNN_model}: 2 $LBconv$ blocks, followed by 2 convolutional layers and 1 $LBconv$ block providing a 2048 feature map (e.i., $[128\times4\times4]$) that is fed to $LBflat$. The output of the latter is provided to the final block, named here $LBFcr$. Finally, a Fully Connected (FC) layer of output $[1\times1]$ is required to generate the quality score. The three sub-networks left, right and stereo extract relevant feature maps from the left, right and stereo images, respectively. We concatenate the outcomes of the three subnetworks (i.e. $FC l$, $FC r$, and $FC s$) and fed them into the $LBconct$ block to form a global feature map of size $[1\times1536]$ used for global prediction. This concatenation allows the individual binocular features to be combined, which serves the model in determining whether the quality is good or not.

The model uses 32 × 32 patches from each view without normalization, clipped consecutively without overlapping from distorted stereoscopic images with a stride of 32 pixels, where each patch has the same score as its associated source. The predicted quality score $Q$ of the whole input stereo image is finally reported by computing the mean of patches scores.

\vspace{-3mm}
\subsection{Model training}
\vspace{-3mm}

In order to minimize the error during the training of the designed end-to-end CNN model, we use the $Loss$ function which is a linear combination of four $L_{1}$ loss functions as described in the following:
\begin{multline}
    Loss = 2 \times |Q_{g} - M_{s}| + |Q_{s} - M_{s}| + \\ |Q_{l} - M_{l}| +|Q_{r} - M_{r}| 
\end{multline}
where $Q_{g}$, $Q_{s}$, $Q_{l}$, and $Q_{r}$ are the predicted quality score of the global, stereo, left and right image, respectively. The $M_{s}$, $M_{l}$ and $M_{r}$ refer to the MOS of the stereo, left, and right image, respectively. 

To update the weights of the whole network, we used the Stochastic Gradient Descent (SGD) with a momentum factor equals to 0.9, a weight decay factor sets to 0.0001, a mini batch size equals to 128 and a learning rate initialized to 10-3. The Pytorch framework was used to implement our approach.
\vspace{-3mm}

\vspace{-3mm}
\section{EXPERIMENTAL RESULTS}
\label{sec:experimental}
\vspace{-2mm}

\subsection{Dataset and training protocol}
\vspace{-3mm}
A database of images with quality scores for each image is required for training and evaluating SIQA measures. The quality score is often obtained by subjective scoring and shown as MOS/DMOS According to ITU-T P.910 \cite{itu1999subjective} recommendations. For the SIQA domain, a variety of databases are publicly available for the implementation and evaluation of the proposed metrics \cite{moorthy2013subjective, chen2013full, wang2014quality, wang2015quality}. However, only Waterloo IVC 3D Phase 1 and Waterloo IVC 3D Phase 2 provide MOS scores for each view, in addition to stereo MOS scores. Moreover, the asymmetric degradations in the Waterloo P-1 and P-2 databases are different from those in the LIVE-II database \cite{chen2013full}. LIVE-II uses only one type of distortion to perform the asymmetry, while the two Waterloo databases consider the possibility of multiple types of degradation in which the left and the right images are affected by different distortions. Therefore, in our validation experiments, Waterloo-P1 and P2 have been used, described as: \textbf{Waterloo IVC 3D Phase 1 (P1) \cite{wang2014quality}} includes 330 full HD (1920 x 1080 pixel) distorted stereo images generated from six pristine stereo pictures through three types of distortion: additive white Gaussian noise, Gaussian blur and JPEG compression. Subjective evaluation scores are expressed in terms of MOS and are distributed in the range [10,100], where 100 indicates the best quality score. \textbf{Waterloo IVC 3D Phase 2 (P2) \cite{wang2015quality}} includes 460 full HD stereo images made up of 10 pristine stereo image pairs by considering the same distortion types. Both datasets comprise symmetric and asymmetric distortions. Subjective assessment scores are in terms of MOS, with the same range as Waterloo-P1 ([10,100]).

The performance of our method was quantified using the two databases. To guarantee that our model evaluates the image quality rather than focusing on the content, we divide each database into 80\% for training and rest 20\% for test based on reference images. So the training images scene are independent from those used in test phase. We repeat the same process 10 times and report the average performance. During the training, we do not consider data augmentation because the noted subjective score may differ if data augmentation techniques are used (e.g., rotation, flipping, etc.) \cite{messai20223d}.
\vspace{-3mm}
\subsection{Comparison with the State-of-the-Art}
\vspace{-3mm}

The performance has been measured across three metrics: The \emph{RMSE}, Pearson linear correlation coefficient (\emph{PLCC}), Spearman's rank order correlation coefficient (\emph{SROCC}) between the machine quality judgments (objective scores) and the human ratings (subjective scores). High values for \emph{PLCC} and \emph{SROCC} (close to 1) and low values for \emph{RMSE} (close to 0) indicate a better prediction performance. 

Overall, the statistical association between human quality scores and our method ratings exhibited outstanding performance and consistency. The obtained results were compared to many FR and NR-SIQA. Among them, there are recent reference-less metrics based on the use of CNN models, namely Chen \cite{chen2020blind} and Sun \cite{sun2020learning}. Table \ref{T:Waterloo-all} shows the results of these methods on both Waterloo-P1 and P2 datasets. Best metric of each category (FR and NR) is represented on bold and the best one whatever the category is with a gray background. As can be seen, our metric outperforms all the state-of-the-art NR and FR metrics on both databases. Furthermore, we report the performance of our method according to the size of the training set. Table \ref{T:DP} shows the correlations achieved for a training set of size 50\%, 70\% and 80\%. The partition ratio has a slight impact on the performance. And it does not suffer from an over-fitting problem. The diminution is similar for both datasets. 
\vspace{-3mm}

\begin{table}[ht]
\centering
\caption{Overall performance comparison on Waterloo-P1 and Waterloo-P2.}
\resizebox{1\columnwidth}{!}{%
\begin{tabular}{c|c|c c c||c c c} 

&  &         &  Waterloo-P1       &   &      &  Waterloo-P2    &   \\ \hline
Type          & Metrics & SROCC        &  PLCC    & RMSE & SROCC        &  PLCC    & RMSE\\  \hline 

&Benoit \cite{benoit2009quality}        &0.332 &0.332 &- &0.165 &0.320 &- \\   

FR&Chen \cite{chen2013full}   &0.457 &0.631 &- &0.272 &0.442 &- \\ 

&Ma \cite{ma2017full}       &\textbf{0.911} &\textbf{0.925} &\textbf{5.876}  &- &-  &-\\

\hline 


 &DECOSINE \cite{yang2018blind} &0.924 &0.943 &- &0.914 &0.933 &- \\

 &Yang \cite{yang2018predicting} & 0.911 &0.940  &-  &0.866  &0.899  &- \\

 &Chen \cite{chen2020blind} &0.923  & 0.931& 5.989 &0.922 &0.925 &7.119 \\

NR  &Sun \cite{sun2020learning} &- &-  &- &0.835 & 0.840&-\\ 
  
  &Liu \cite{liu2020no} &0.928 &0.945  &5.268 &0.901 &0.913 &7.658 \\ 
  
  &Wang \cite{wang2021no} &0.950 &0.959  &4.089 &0.953 &0.965 &4.498 \\ 
 
\rowcolor{Gray} &  Proposed   &\textbf{0.967} &\textbf{0.972} &\textbf{3.635} &\textbf{0.966} &\textbf{0.971}  &\textbf{4.161}  \\ \hline 
\end{tabular}
\label{T:Waterloo-all}
}
\end{table}
\vspace{-3mm}
\vspace{-4mm}
\begin{table}[h]

\caption{Performance of the proposed metric under different train-test partitions on Waterloo-P1 and Waterloo-P2.}
\resizebox{1\columnwidth}{!}{%
\begin{tabular}{c|c c c||c c c}
       &        &Waterloo-P1    &      &      &Waterloo-P2    \\ 
\hline
Partition & SROCC        &  PLCC    & RMSE & SROCC        &  PLCC    & RMSE\\  
\hline 

80\%-20\% &\textbf{0.967} &\textbf{0.972} &\textbf{3.635} &\textbf{0.966} &\textbf{0.971}  &\textbf{4.161}   \\ 

70\%-30\% &0.961 &0.968 &3.876 &0.952 &0.962  &4.964   \\ 

50\%-50\%   &0.947 &0.964 &4.276 &0.944 &0.958  &6.226   \\ 

\hline 

\end{tabular}
} 
\label{T:DP}
\end{table}
\vspace{-3mm}
\vspace{-4mm}
\vspace{-2mm}
\subsection{Cross database performance}
\vspace{-3mm}

Cross-database experiments have been conducted in order to verify the generalization ability of the proposed approach. The implemented tests are shown in Table \ref{T:cross-performance}. Metrics shown are all NR methods. They have been trained in the former database and tested on the latter. Comparing with the NR metrics, our method has outstanding quality prediction performance in terms of \emph{PLCC}. Saliency-SIQA, DECOSINE and Wang algorithms deliver decent performance, but the proposed one is the only metric which gives performance over 0.9. Compared to the second best metric (i.e., Wang), the improvement accuracy of quality assessment was found to be 10\% in terms of \emph{PLCC}.

\begin{table}[t]
\centering

\caption{PLCC Performance of cross-database tests using the two databases (Expressed as: Train database/Test database).}
\resizebox{1\columnwidth}{!}{%
\begin{tabular}{c | c | c}

 Metrics    &  Waterloo-P1/Waterloo-P2 &Waterloo-P2/Waterloo-P1    \\ 
\hline 

Liu \cite{liu2020no} &0.696 &0.701  \\

Yang \cite{yang2018predicting} & 0.781 &0.864  \\

Chen \cite{chen2020blind}& 0.806 &0.846  \\

Saliency-SIQA \cite{messai20223d} &  0.826 &  0.848  \\

DECOSINE \cite{yang2018blind}  & 0.842 & 0.873  \\

Wang \cite{wang2021no} & 0.856 &0.881  \\

Proposed &  \textbf{0.944} &  \textbf{0.940}  \\

\hline 

\end{tabular}
} 
\label{T:cross-performance}
\end{table}

\vspace{-3mm}
\subsection{Ablation study and run-time}
\vspace{-3mm}

In the ablation test case, we simply erase the loss of left and right views. As a result, the model is entirely trained on stereo MOS loss. Table \ref{T:ablation_study} shows the performance without using left and right human ratings versus using them. Furthermore, we investigated the advantage of utilizing the global score as the final quality score rather than the stereo score. The results show that our concept improves performance and support the idea of a multi-score model.
\vspace{-3mm}

\begin{table}[h]
\caption{Performance obtained of ablation tests on Waterloo-P1 and Waterloo-P2.}
\resizebox{1\columnwidth}{!}{%
\begin{tabular}{c|c c c||c c c}
       &        &Waterloo-P1    &      &      &Waterloo-P2    \\ 
\hline
Model & SROCC        &  PLCC    & RMSE & SROCC        &  PLCC    & RMSE\\  
\hline 

Without global score &0.924 &0.946 &5.480 &0.936 &0.942  &5.703  \\ 

Without left, Right MOS &0.949 &0.962 &4.965 &0.955 &0.966  &4.562  \\ 

\textbf{Proposed}: With left, Right MOS &\textbf{0.967} &\textbf{0.972} &\textbf{3.635} &\textbf{0.966} &\textbf{0.971}  &\textbf{4.161}\\ 
   
\hline 
\end{tabular}
} 
\label{T:ablation_study}
\end{table}
\vspace{-3mm}

We measured less than 2 seconds of run-time (\textbf{1.485 s}) using a single Full-HD stereoscopic image (1920 x 1080 pixels), including patch cropping and loading using a Dell Precision 7550 laptop equipped with an Intel i9-10885H CPU @ 2.40GHz processor and an NVIDIA Quadro RTX 3000 GPU. In terms of run-time speed, the proposed metric has the potential to be used on stereoscopic videos.

\vspace{-3mm}
\section{CONCLUSION}
\label{sec:conclusion}
\vspace{-3mm}

In this paper, we introduced a new NR-SIQA approach based on a multi-score convolutional neural network model for evaluating the quality of left, right and global stereoscopic images. We used the corresponding MOS for each view to learn the best quality indicators task to improve the quality prediction. Based on a comparative examination using two public databases, Waterloo-P1 and Waterloo-P2, our model outperforms state-of-the-art approaches, especially in cross-validation tests. The ablation study demonstrates that using a multi-score task is a promising path for improving the accuracy of quality score prediction. In addition, we employ RGB color channels rather than gray ones to maintain the same spatial domain being viewed by the observer. As future work, we will undertake a detailed examination and consider extending the model to handle auxiliary tasks such as distortion type recognition and degree of deformations.


\bibliographystyle{IEEEbib}
\bibliography{strings,refs}

\end{document}